\documentclass[letterpaper, paper,10pt]{AAS}		

\usepackage{bm}
\usepackage{array}
\usepackage{amsmath}
\usepackage{amssymb}
\usepackage{subfigure}
\usepackage[colorlinks=true, pdfstartview=FitV, linkcolor=black, citecolor= black, urlcolor= black]{hyperref}
\usepackage{overcite}
\usepackage{footnpag}			      	
\usepackage{float}
\usepackage{comment}

\PaperNumber{19-628}

\begin{document}

\title{Seeker based Adaptive Guidance via Reinforcement Meta-Learning Applied to Asteroid Close Proximity Operations}
\author{Brian Gaudet\thanks{Co-Founder, DeepAnalytX LLC, 1130 Swall Meadows Rd, Bishop CA 93514 and Affiliated Engineer Department of Systems and Industrial Engineering, University of Arizona, 1127 E. Roger Way, Tucson Arizona, 85721 briangaudet@mac.com},\hspace{0.1cm}  
Richard Linares\thanks{Charles Stark Draper Assistant Professor, Department of Aeronautics and Astronautics, Massachusetts Institute of Technology, Cambridge, MA 02139},
\ and Roberto Furfaro\thanks{Professor, Department of Systems and Industrial Engineering, Department of Aerospace and Mechanical Engineering, University of Arizona, 1127 E. Roger Way, Tucson Arizona, 85721}
}

\maketitle{}

\begin{abstract}
Current practice for asteroid close proximity maneuvers requires extremely accurate characterization of the environmental dynamics and precise spacecraft positioning prior to the maneuver. This creates a delay of several months between the spacecraft's arrival and the ability to safely complete close proximity maneuvers. In this work we develop an adaptive integrated guidance, navigation, and control system that can complete these maneuvers in environments with unknown dynamics, with initial conditions spanning a large deployment region, and without a shape model of the asteroid. The system is implemented as a policy optimized using reinforcement meta-learning.  The spacecraft is equipped with an optical seeker that locks to either a terrain feature, back-scattered light from a targeting laser, or an active beacon, and the policy maps observations consisting of seeker angles and LIDAR range readings directly to engine thrust commands. The policy implements a  recurrent network layer that allows the deployed policy to adapt real time to both environmental forces acting on the agent and internal disturbances such as actuator failure and center of mass variation. We validate the guidance system  through simulated landing maneuvers in a six degrees-of-freedom simulator. The simulator randomizes the asteroid's characteristics such as solar radiation pressure, density, spin rate, and nutation angle, requiring the guidance and control system  to adapt to the environment. We also demonstrate robustness to actuator failure, sensor bias, and changes in the spacecraft's center of mass and inertia tensor. Finally, we suggest a concept of operations for asteroid close proximity maneuvers that is compatible with the guidance system. 

\end{abstract}

\section{Introduction}

Current practice for asteroid close proximity operations involves first making several passes past the asteroid in order to collect images and LIDAR data that allows creating a shape model of the asteroid \cite{gaskell2008characterizing}. In addition, these maneuvers can be used to estimate the environmental dynamics in the vicinity of the asteroid, which allows calculation of a burn that will put the spacecraft into a safe orbit \cite{udrea2012sensitivity}. Once in this safe orbit, statistical orbit determination techniques \cite{schutz2004statistical} augmented by optical navigation are used to create a model that can estimate the spacecraft’s orbital state (position and velocity) as a function of time. This model is extremely accurate provided the spacecraft remains in the safe orbit. While the spacecraft is in the safe orbit, both the asteroid shape model and the model of the asteroid’s dynamics are refined. Mission planners then use the orbit model, along with an estimate of the forces acting on the spacecraft, to plan an open loop maneuver that will bring the spacecraft from a given point in the safe orbit to some desired location and velocity. An example of such a trajectory is the Osiris Rex Touch and Go (TAG) sample collection maneuver \cite{udrea2012sensitivity}. If the dynamics model used to plan the open loop maneuver is not completely accurate, errors can accumulate over the trajectory, resulting in a large error ellipse for the contact position between the asteroid and spacecraft. For this reason, multiple test maneuvers are typically run in order to insure the dynamics model is accurate in the region of the maneuver. Finally, note that current practice is not compatible with completely autonomous missions, where the spacecraft can conduct operations on and around an asteroid without human supervision.

Now consider an adaptive guidance, navigation and control (GNC) system that after a short interaction with the environment can adapt to that environment's ground truth dynamics, only limited by the spacecraft's thruster capability.  Such a system would allow a paradigm shift for mission design, as highlighted in the comparison between current practice and what might be possible using the proposed system, as shown in Table \ref{tab:innovation}. Of course there are scientific reasons for characterizing an asteroid's environment as accurately as possible, but the proposed innovation gives mission flexibility.  For example, a mission might involve visiting multiple asteroids and collecting samples, and the orbits of the asteroids might make it necessary to spend only a short time at each one. Or the mission goal might not be scientific at all, but rather to identify resource rich asteroids for future mining operations. For a given level of accuracy with respect to the environmental dynamics model, the ability to adapt real time when the environment diverges from the model should provide a significant reduction in mission risk.

\begin{table}[!ht]
	\fontsize{10}{10}\selectfont
    \caption{A Comparison of Current Practice and Proposed Concept of Operations.}
   \label{tab:innovation}
        \centering 
   \renewcommand{\arraystretch}{1.5}
   \begin{tabular}{p{4.6cm} | p{4.6cm} | p{4.6cm} } 
      \bf{Metric} & \bf{Current Practice} & \bf{Adaptive GNC System}\\
      \hline
      Dynamics Characterization & Necessary to characterize dynamics to a very high degree of accuracy & Dynamics must be characterized to the extent required for thruster sizing \\
      Planning & Requires time and fuel consuming trajectory planning, usually requiring multiple test maneuvers followed by telemetry analysis & Once a landing site has been identified (see Concept of Operations) the spacecraft's guidance, navigation, and control system can autonomously collect the sample \\
      Initial Conditions & Requires the spacecraft to begin at  a specific initial condition (sub meter position and sub cm/s velocity knowledge) & Deployment ellipse can span several cubic km, and large initial heading and attitude errors can be tolerated \\
   \end{tabular}
\end{table}

Coupling a suitable navigation system with a traditional closed loop guidance and control law (such as DR/DV \cite{d1997optimal}) can potentially improve maneuver accuracy. However, if the asteroid's environmental dynamics are not well characterized, accuracy will still be compromised due to errors stemming from both the dynamics model used in the state estimation algorithm and the potential inability of the guidance and control law to function optimally in an environment with unknown dynamics. Indeed, an optimal trajectory generated based off of an inaccurate dynamics model may be infeasible (impossible to track with a controller given control constraints) in the actual environment. Moreover, our initial research into this area \cite{gaudet2019adaptive} has shown that traditional closed loop guidance laws such as DR/DV \cite{d1997optimal} are not robust to actuator failure, unknown dynamics, and navigation system errors, whereas the proposed GNC system is.  Finally, note that integration of the navigation system allows the system to quickly adapt to sensor bias.

Recent work by others in the area of adaptive guidance algorithms include Reference (\citenum{prabhakar2018trajectory}), which demonstrates an adaptive control law for a UAV tracking a reference trajectory, where the adaptive controller adapts to external disturbances. One limitation is the linear dynamics model, which may not be accurate, as well as the fact that the frequency of the disturbance must be known. Reference (\citenum{huang2019mars}) develops a fault identification system for Mars entry phase control using a pre-trained neural network, with a fault controller implemented as a second Gaussian neural network,  Importantly, the second network requires on-line parameter update during the entry phase, which may not be possible to implement in real time on a flight computer. Moreover, the adaptation is limited to known actuator faults as identified by the 1st network. Reference (\citenum{han2015adaptive}) develops an adaptive controller for spacecraft attitude control using reaction wheels. This approach is also limited to actuator faults, and the architecture does not adapt to either state estimation bias or environmental dynamics.  

In this work we develop an adaptive and integrated GNC system applicable to asteroid close proximity maneuvers. The  system is optimized using reinforcement meta-learning (RL meta-learning), and implements a global policy over the region of state space defined by the deployment region and potential landing sites. Reinforcement Learning (RL) has recently been successfully applied to landing guidance problems \cite{furfaro2018deep,furfaro2018recurrent,furfaro2017waypoint3,gaudet2018deep}. Importantly, the observations are chosen such that the policy generalizes well to different landing sites. Specifically, the policy can be optimized for a specific landing site, and when deployed can be used for an arbitrary landing site. The policy maps observations to actions, with the observations consisting of angles and range readings from the spacecraft's seeker, changes in spacecraft attitude since the start of the maneuver, and spacecraft rotational velocity. The policy actions consist of on/off thrust commands to the spacecraft's thrusters.  In order to reduce mission risk, we present a concept of operations (CONOPS) that tags the landing site with a targeting laser, providing an obvious target for the seeker's camera. However, future work will investigate the effectiveness of using terrain features as targets. In the RL framework, the seeker can be considered an attention mechanism, determining what object in the agent's field of regard the policy should target during the maneuver. In the case where we want to target a terrain feature rather than a tagged landing site, the landing site would be identified by the seeker, rather than the guidance policy. Both seeker design and laser aided guidance are mature technologies, with seekers being widely used in guided missiles \cite{siouris2004missile}, and laser aided guidance used in certain types of missiles and guided bombs. 


Adaptability is achieved through RL-Meta Learning, where different environmental dynamics, sensor noise, actuator failure, and changes in the spacecraft's center of mass and inertia tensor are treated as a range of partially observable Markov decision processes (POMDP). In each POMPD, the policy's recurrent network hidden state will evolve differently over the course of an episode, capturing information regarding hidden variables that are useful in minimizing the cost function, i.e.,  external forces, changes in the spacecraft's internal dynamics and sensor bias.  By optimizing the policy over this range of POMDPs, the trained policy will be able to adapt to novel POMPDs encountered during deployment. Specifically, even though the policy's parameters are fixed after optimization, the policy's hidden state will evolve based off the current POMPD, thus adapting to the environment.

\begin{figure}[h]
\begin{center}
\includegraphics[width=.9\linewidth]{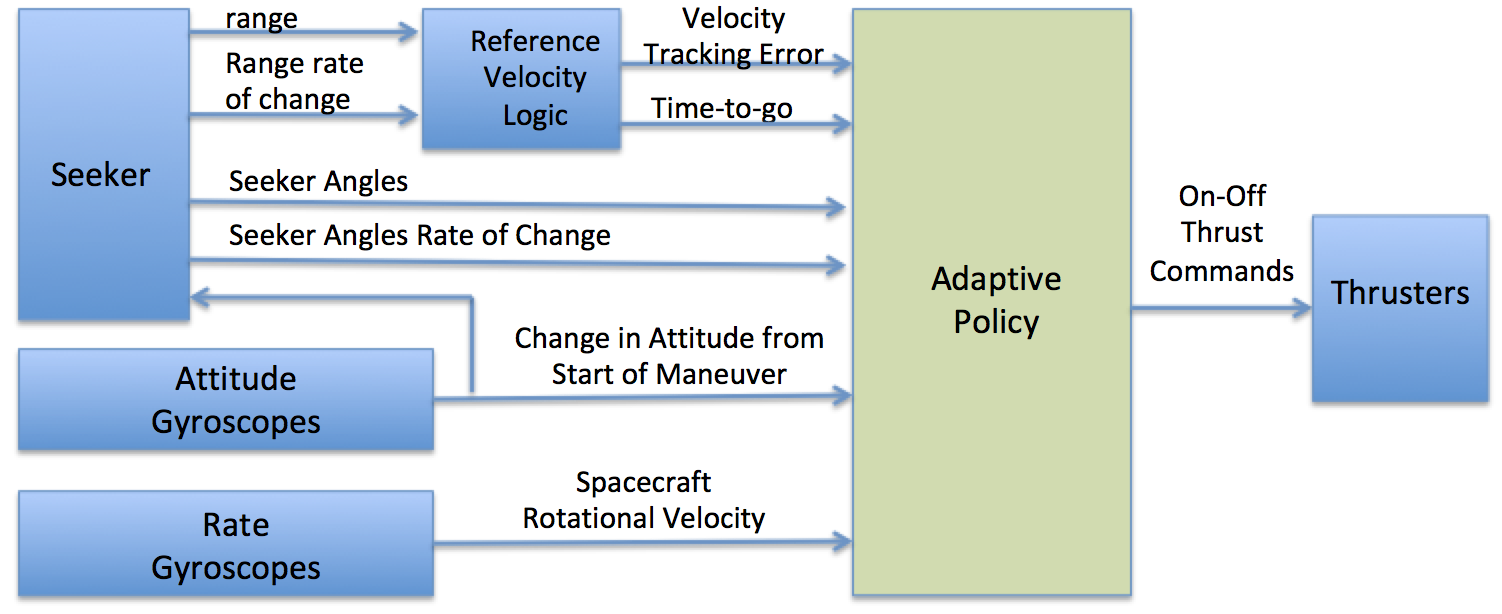}
\caption{GNC Policy and Peripheral Spacecraft Components}
\label{fig:system}
\end{center}
\end{figure}

The policy is expected to use approximately 16,000 32 bit network parameters, and will require approximately 64KB of memory. In our previous work with 6-DOF Mars powered descent phase the policy took less than 1mS to run the mapping between estimated state and thruster commands (four small matrix multiplications) on a 3Ghz processor\cite{gaudet2018deep}. Since in this work the mapping is updated every six seconds, we do not see any issues with running this on the current generation of space-certified flight computers. A diagram illustrating how the policy interfaces with peripheral spacecraft components is shown in Fig.~\ref{fig:system}.

One advantage of our proposed CONOPS and GN\&C system as compared to current practice is that the environmental dynamics need not be accurately characterized prior to the maneuver, removing an element of mission risk. Compared to completely passive optical navigation approaches, our method has the additional advantage that it is insensitive to lighting conditions and does not rely on the asteroid having sufficient terrain diversity to enable navigation. Moreover, the system can adapt to sensor bias and actuator failure, further reducing mission risk. The downside is that fuel efficiency will be inferior to that of an optimal trajectory generated using knowledge of the environmental dynamics. It would be possible to improve the fuel efficiency if after observing the movement of the target location from an inertial reference frame, the spacecraft could be put on a collision triangle heading with the target landing site. Instead of heading for the target site, the lander would head towards the  point where the target site will be at the completion of the maneuver. In this approach, the agent would be rewarded for keeping the seeker angles at their value at the start of a maneuver, which will keep the spacecraft on the collision triangle with the moving target, as described in more detail in (Reference \citenum{2019arXiv190602113G}).

We  demonstrate that the system can complete maneuvers from a large deployment region and without knowledge of the local environmental dynamics, and successfully adapt to sensor distortion, changes in the spacecraft's center of mass and inertia tensor,  and actuator failure. In this work, we will focus on a maneuver that begins approximately 1km from the desired landing site, with a deployment region spanning approximately 1 cubic km. The goal is to reach a position within 1m of a target location 10m above the designated landing site, with velocity magnitude less than 10cm/s, and negligible rotational velocity.  What happens next will be mission specific. To illustrate a scenario, a hovering guidance and control system using LIDAR altimeter could take over at that point, bring the spacecraft to an attitude consistent with the deployment of a robotic arm, and collect a sample, with the hovering controller compensating for the disturbance created by the arm pushing against the surface.  Alternately, the spacecraft could release a rover from this altitude. 

\section{Concept of Operations}

Once the spacecraft arrives at the target asteroid the spacecraft will hover in proximity to the asteroid using LIDAR altimetry.  Next, the spacecraft illuminates a point on the asteroid's surface with its targeting laser and deploys a cubesat, which uses the proposed guidance system to land at the tagged location. At that point the cubesat activates its beacon, which emits light at a wavelength that gives good contrast with the ambient lighting. The process is repeated in multiple locations so that at any hovering point above the asteroid, the spacecraft has line of sight to one of the beacons, each with a beacon emitting at a different wavelength. The spacecraft's seeker can then lock onto one of the beacons.  Given the seeker platform's attitude (which is stabilized during these operations), the seeker range and gimbal angles are used to determine the spacecraft's location in a beacon centered reference frame. Remote sensing measurements can then be stored in a map that contains (with attitude potentially defined in the celestial reference frame):

\begin{enumerate}
\item The beacon associated with the remote sensing measurement
\item The seeker platform's stabilized attitude at the time of measurement (required to infer position from range and seeker angles)
\item The spacecraft's position in that beacon's reference frame
\item The remote sensing platform's attitude at the time of measurement
\end{enumerate}

When planetary geologists have decided on an area of high scientific or economic interest, the map allows the spacecraft to easily navigate back to a position that gives line of sight to the area of interest and set the remote sensing platform's attitude to the value at which the measurement was taken. This allows the targeting laser to be pointed in a direction that tags the area of interest.  Once the area of interest is illuminated with the targeting laser,  two scenarios are possible: 

\begin{enumerate}
\item \underline{Cubesat Sample Collection:} Here a cubesat conducts a maneuver using the proposed GNC system that terminates with the cubesat hovering above the targeted location. The cubesat can then hover at various locations close to the beacon and collect samples, using autonomous hazard detection software to avoid collisions with scarps or boulders.  Once the samples are collected, the spacecraft turns on an optical homing beacon, allowing the cubesat to rendezvous with the spacecraft using the proposed GNC system and return the samples to the spacecraft. An obvious advantage of this approach is that if something goes wrong, the spacecraft can launch a 2nd cubesat and try again.

\item \underline{Spacecraft Sample Collection:}   
Here a cubesat is deployed to land at the the specified location.  Once in place, this cubesat activates its beacon, allowing the spacecraft to navigate to within a range that allows deployment of sample collection manipulators.  The spacecraft can then hover at various locations close to the beacon and collect samples, using autonomous hazard detection software to avoid collisions with scarps or boulders.  
\end{enumerate}

The proposed guidance system is not limited to landing maneuvers. Given an established beacon centered reference frame, it can also be used to move between high altitude positions that have line of sight to the given beacon.  Specifically, for  navigation between hovering  positions in a beacon centered reference frame, the actual seeker locks to the reference beacon, and the seeker platform's attitude and seeker angles are used to infer position in the beacon centered reference frame.  Given a new target hovering location in the same reference frame, the guidance system then simulates the measurement from a "virtual seeker" locked onto the target hovering location, allowing the spacecraft to navigate to the new hovering position. In contrast, for landing maneuvers, the seeker locks onto either the back-scattered light from the targeting laser (cubesat Sample Collection) or the landing site beacon (Spacecraft Sample Collection), and the guidance system then brings the spacecraft to rest 10m above the desired landing site. Note that what we refer to as a cubesat could be larger than the standard cubesat dimensions if mission requirements dictate a larger vehicle. Although cubesat dimensions should suffice for a beacon, somewhat larger dimensions (say 20cm x 20cm) might be required for sample return hardware. Also note that there are several off the shelf propulsion systems that could be used in this application (i.e., the lunar flashlight system from Vacco). Finally, note that the navigation camera and LIDAR altimeter can be small and low quality, as the proposed system is robust to sensor noise.

Although in this CONOPS a shape model is not necessary, the navigation beacons will provide reference frames that should aid in shape model generation. Specifically, rather than require full simultaneous location and mapping \cite{thrun2005probabilistic} (SLAM), the mapping can occur in a reference frame anchored to points on the asteroid surface.  Also note that while it might seem extravagant to deploy several cubesats, their volume and mass is minimal. Moreover, although in theory it is possible to navigate using only terrain features, such features tend to be obvious from high altitudes but less so when the spacecraft approaches the surface. In contrast, the proposed CONOPS provides a simple and robust framework for asteroid close proximity operations.  And if the deployed cubesats have a sleep mode to conserve power, they can potentially be reused for return missions to the asteroid. Another advantage of the proposed CONOPS is that the sample collection can occur shortly after the spacecraft arrives at the asteroid, with most of the time between arrival and sample collection consisting of collecting remote sensing measurements and analyzing them on Earth for landing site selection.   If landing site selection were to be automated, fully autonomous exploration would be possible.  The use of cubesats opens many mission possibilities. To illustrate, rather than tediously drilling through the asteroid's surface to obtain a sub-surface sample, a cubesat could carry a payload of several kg of C4 explosive, with the contact fuse armed after launch from the spacecraft, and a targeted terminal velocity of 0.5m/s. This would likely create a small crater, after which samples could be collected from the crater bottom. This paper focused on the integrated GNC system that allows pinpoint landing at a site illuminated with a targeting laser or landing beacon, and leaves the development of the hovering controller, navigation between hovering positions, and hazard avoidance software to future work. In the sequel, the term "beacon" will refer to either an active light emitting beacon or back-scattered light from a targeting laser.

\section{Problem Formulation}

\subsection{Spacecraft Configuration}

The spacecraft is modeled as a uniform density cube with height $h=2\text{m}$, width $w=2\text{m}$, and depth $d=2\text{m}$, with inertia matrix given in Eq.~\eqref{eq:inertia_tensor}, where $m$ is the spacecraft's mass. The spacecraft has a wet mass ranging from 450 to 500 kg. The thruster configuration is shown in Table~\ref{tab:thrusters}, where $x$, $y$, and $z$ are the body frame axes.  Roll is about the $x$-axis, yaw is about the $z$-axis, and pitch is about the $y$-axis. When two thrusters on a given side of the cube are fired concurrently, they provide translational thrust without rotation. Firing offset thrusters on two opposite sides provides rotation without translation. In general, the configuration gives full translational and rotational motion, and is robust to the failure of any single thruster. 

\begin{equation}
    \label{eq:inertia_tensor}
    {\bf J}=\frac{m}{12}\begin{bmatrix} h^2 + d^2 & 0 & 0 \\ 0 & w^2+d^2 & 0 \\ 0 & 0 & w^2+h^2\end{bmatrix}
\end{equation}

\begin{table}[!ht]
	\fontsize{10}{10}\selectfont
    \caption{Body Frame Thruster Locations.}
   \label{tab:thrusters}
        \centering 
   \begin{tabular}{c | r | r | r | r | l} 
      \hline
      Thruster & x (m) & y (m) & z (m) & Thrust & Description\\
      \hline
       1 & -1.0 & 0.0 & 0.4 & 1.0 & translational and pitch stabilizer  \\
      2 & -1.0 &  0.0 & -0.4 & 1.0 & translational and pitch stabilizer \\
      3 & 1.0 & 0.0 & 0.4  & 1.0 & translational and pitch stabilizer \\
      4 & 1.0 & 0.0 & -0.4 & 1.0 & translational and pitch stabilizer \\
      5 & -0.4 & -1.0 & 0.0 & 1.0 & translational and yaw stabilizer \\
      6 & 0.4 & -1.0 & 0.0 & 1.0 & translational and yaw stabilizer \\
      7 & -0.4 & 1.0 & 0.0 & 1.0 & translational and yaw stabilizer \\
      8 & 0.4 & 1.0 & 0.0 & 1.0 &  translational and yaw stabilizer \\
      9 & 0.0 & -0.4 & -1.0 & 1.0 & translational and roll stabilizer \\
      10 & 0.0 & 0.4 & -1.0 & 1.0 & translational and roll stabilizer \\
      11 & 0.0 & -0.4 & 1.0 & 1.0 & translational and roll stabilizer \\
      12 & 0.0 & 0.4 & 1.0 & 1.0 & translational and roll stabilizer \\
   \end{tabular}
\end{table}

We assume that the spacecraft is equipped with a stabilized seeker with ideal tracking capability, with a 90 degree field of regard, mounted on the side with the -Z surface normal. The seeker consists of a camera and LIDAR altimeter that point along the same direction vector (boresight axis). At the start of a maneuver, the seeker platform's attitude is aligned with the spacecraft body frame such that the seeker boresight axis is aligned with the -Z surface normal. The seeker is mechanically controlled to adjust the seeker angles (elevation and azimuth) between the boresight axis and the seeker platform such that the target is centered in the camera field of view. Consequently, when the seeker is locked onto the target, the LIDAR altimeter gives the range to target. Further, the seeker platform is stabilized so that seeker measurements are taken from an inertial reference frame. Stabilization can be accomplished by measuring changes in the spacecraft's attitude during the maneuver and mechanically adjusting the seeker platform's attitude so that it remains constant for the duration of the maneuver.  Stabilization prevents changes in the spacecraft's attitude during a maneuver being interpreted as a change in target position.

As the seeker tracks the target from this stabilized reference frame, we can define the angles between the seeker boresight axis and the seeker reference frame $x$ and $y$ axes as the seeker angles $\theta_u$ and $\theta_v$.  Further, define $\mathbf{C}_\mathrm{SN}(\mathbf{q}_0)$ as the direction cosine matrix (DCM) mapping from the inertial frame to the stabilized seeker platform reference frame, with $\mathbf{q}_0$ being the spacecraft's attitude at the start of the maneuver. We can now transform  the target's relative position in the inertial reference frame $\mathbf{r}_\mathrm{TM}^\mathrm{N}$ into the seeker reference frame as shown in Eq.~\eqref{eq:seeker1}.

\begin{equation}
\label{eq:seeker1}
\mathbf{r}_\mathrm{TM}^\mathrm{S} = [\mathbf{C}_\mathrm{SN}(\mathbf{q}_0)]\mathbf{r}_\mathrm{TM}^\mathrm{N}
\end{equation}

Defining the line of sight unit vector in the seeker reference frame as $\boldsymbol{\hat\lambda}^\mathrm{S} = \dfrac{\mathbf{r}_\mathrm{TM}^\mathrm{S}}{\|\mathbf{r}_\mathrm{TM}\|^\mathrm{S}}$ and the seeker frame unit vectors $u=\begin{bmatrix} 1 & 0 & 0 \end{bmatrix}$ , $v=\begin{bmatrix} 0 & 1 & 0\end{bmatrix}$  we can then compute the seeker angles as the orthogonal projection of the seeker frame LOS vector onto $u$ and $v$ as shown in Eqs.~\eqref{eq:seeker3} and \eqref{eq:seeker4}.

\begin{subequations}
\begin{align}
\theta_{u} &= \mathrm{arcsin}(\boldsymbol{\hat\lambda}^\mathrm{S} \cdot \hat{u})\label{eq:seeker3}\\
\theta_{v} &= \mathrm{arcsin}(\boldsymbol{\hat\lambda}^\mathrm{S} \cdot \hat{v})\label{eq:seeker4}
\end{align}
\end{subequations}

Recall that the seeker has a LIDAR altimeter aligned with the seeker boresight axis. The integrated GNC policy described in the following will map these seeker angles $\theta_u$ and $\theta_v$, their rate of change, as well as the measured range to target and its rate of change to thrust commands. Since it is possible to sample from the seeker at a much higher frequency than the guidance frequency, it should be possible to obtain accurate seeker angle and time derivative ($\dot\theta_u$ and $\dot\theta_v$) measurements by averaging the seeker angles over the guidance period.

\subsection{Landing Scenario}

The initial condition limits were selected based off of the expected CONOPS, where the spacecraft would begin from a hovering position with line of sight to the target landing site.  From this position, the seeker would lock onto the target beacon, and the spacecraft would rotate so that the spacecraft's -Z axis is aligned with the line of sight, i.e., the seeker angles would be zero with the seeker locked onto the beacon. At this point the seeker platform's attitude is reset so that the seeker boresight axis is aligned with the spacecraft's -Z axis. From here a short open loop burn is initiated for 10s in order to achieve a positive closing velocity which avoid a singularity in the calculation of the reference speed. During this time, the asteroid can rotate, and the spacecraft's velocity vector and attitude at the end of the burn will no longer be aligned with the line of sight to target at the end of the burn.

The spacecraft will begin at at a random position specified by a distance between 0.8 and 1.0km from the target location, with an offset angle between 0 and 45 degrees. This offset angle is given by $\arccos(\mathbf{\hat{r}}_{ts}\cdot\mathbf{\hat{r}}_{at}$), where $\mathbf{\hat{r}}_{ts}$ is the direction vector pointing from the target to the spacecraft and $\mathbf{\hat{r}}_{at}$ is the direction vector pointing from the asteroid body frame origin to the target. This gives a deployment region of a little over 1 $\text{km}^3$, considerably larger than the sub-meter accuracy required by current practice. Note that an offset angle of zero implies that the asteroid center, target location, and spacecraft are collinear. Defining an optimal heading as the spacecraft's velocity vector $\mathbf{v}$ aligned with $-\mathbf{\hat{r}}_{ts}$, the initial spacecraft velocity will have a random heading error between 0.0 and 22.5 degrees, with a velocity magnitude between 5 and 10 cm/s. The heading error is calculated as $\arccos(\mathbf{-\hat{r}}_{ts}\cdot\mathbf{\hat{v}}$). The spacecraft's ideal initial attitude is such that the -Z body-frame axis is aligned with the line of sight to target. This ideal initial attitude is perturbed at the start of each episode such that the angle between the -Z body frame axis and line of sight to target varies uniformly between 0.0 degrees and 11.3 degrees.  To avoid confusion, note that the angles in Table \ref{tab:initial_conditions} have no preferred direction, i.e.,  they are oriented randomly in $\mathbb{R}^3$. Specifically, we rotate the optimal direction vector for these cases so that it is aligned with the direction vector $\begin{bmatrix} 0 & 0 & 1 \end{bmatrix}$, then we randomly generate the angle $\theta$ over the ranges given in the table, and the angle $\phi$ is randomly set to between -180 and 180 degrees. This gives a random direction vector in the new orientation, which is then rotated back to the original coordinate system. These initial conditions are summarized in Table \ref{tab:initial_conditions}.

\begin{table}[h]
	\fontsize{10}{10}\selectfont
    \caption{Initial Conditions}
   \label{tab:initial_conditions}
        \centering 
   \newcolumntype{R}{>{\raggedleft\arraybackslash}p{4cm}}
   \begin{tabular}{l | R | R } 
        Parameter & min & max \\
       \hline
       Range (m) & 800.0 & 1000.0 \\
      Offset Angle (degrees) & 0.0 & 45.0  \\
      Heading Error (degrees) & 0.0 & 22.5  \\
      Attitude Error (degrees) & 0.0 & 11.3 \\
      Rotational Velocity (mrad/s) & -50.0 & 50.0 \\
   \end{tabular}
\end{table}

We model the asteroid as an ellipsoid with uniform density. We assume that the asteroid is in general not rotating about a principal axis, and therefore to calculate the angular velocity vector we must specify the spin rate, the nutation angle (angle between the asteroid's z-axis and the axis of rotation), and moments of inertia \cite{scheeres2016orbital}.  The moments of inertia in turn depend on the asteroid's density and dimensions. The dimensions are specified by the ellipsoid axes $a>b==c$, where assuming $b==c$ significantly simplifies the equations of motion.  We use a gravity model that assumes a uniformly distributed ellipsoid\cite{scheeres2016orbital}.

\begin{table}[h]
	\fontsize{10}{10}\selectfont
    \caption{Parameters for Randomly Generated Asteroids}
   \label{tab:ast_param}
        \centering 
   \newcolumntype{R}{>{\raggedleft\arraybackslash}p{4cm}}
   \begin{tabular}{l | R | R } 
        Parameter & min & max \\
       \hline
      c (m) & 150 & 300  \\
      b (m) &  c &  c  \\
      a (m) &  b  &  300 \\
      Density $\rho \text{ kg/m}^3$ & 500 & 5000 \\
      Spin Rate $\omega_o$ (rad/s) & $1\times10^{-6}$ & $5\times10^{-4}$   \\
      Nutation Angle (degrees) & 45 & 90 \\
      Acceleration due to SRP  $\text{m/s}^2$ &  $[-100,-100,-100]\times10^{-6}$  &  $[100,100,100]\times10^{-6}$
   \end{tabular}
\end{table}

The adaptability of the system is demonstrated via Monte Carlo simulation by executing the maneuvers under the following  conditions:
\begin{enumerate}
    \item \underline{Unknown Dynamics:} At the start of each episode, the asteroid's spin rate, nutation angle, density, local solar radiation pressure, and axes lengths will be randomly set over the range shown in Table \ref{tab:ast_param}.
    \item \underline{Actuator Failure:} At the start of each episode, with probability $p_{fail}$, a random thruster is selected for failure.  The failure mode is that the thrust gets scaled to between $f_{min}$ and $f_{max}$ of nominal for the duration of the episode. An actuator failure makes it impossible for the spacecraft to decouple translational and rotational motion, reduces maximum translational thrust in a given direction, and reduces torque in a given direction around one of the axes, significantly complicating the guidance and control problem.
    \item \underline{Sensor Distortion:} The seeker's range reading will be biased from the ground truth by multiplying the ground truth range between the spacecraft and target by $1+r_{bias}$. The seeker angle's will be corrupted using a simple refraction model where the measured angle is set to the ground truth angle multiplied by $1 + a_{bias}$. Further, the attitude and angular velocity measurements are biased via multiplication by $1+q_{bias}$ and $1+\omega_{bias}$ respectively. $r_{bias}$, $a_{bias}$, and each element of $q_{bias}$ and $\omega_{bias}$ are drawn uniformly at the start of each episode, and held constant for the duration of the episode. Finally, we add zero mean Gaussian angle noise  with standard deviation ($a_\sigma$) of 1 mrad to the seeker angle measurements. Since current seeker implementations can sample angles at 1 MHz, and our guidance period is greater than 1 second, we could expect that averaging $1\times10^6$ measurements with a standard deviation of 1 rad would give accuracy on the order of 1 mrad (by the central value theorem).
    \item \underline{Inertia Tensor Variation:} We use a simplified model of fuel flow within the spacecraft, so that as the maneuver progresses and fuel is consumed, the translational thrust is not aligned with the center of mass and the inertia tensor varies. This will induce a change in rotational velocity that the GNC system must compensate for. 
    \item \underline{Initial Center of Mass Variation:}  In addition to modeling the dynamic shift of the spacecraft's center of mass due to fuel consumption, at the start of each episode we also perturb the spacecraft's initial center of mass from its nominal position at the center of the spacecraft. This represents uncertainty stemming from fuel consumption prior to maneuver, samples collected by spacecraft, or possibly a cubesat deployed by the spacecraft.
    \item \underline{Initial Mass Variation:}  The spacecraft's initial mass will be uniformly drawn from between 450 and 500kg at the start of each episode.
\end{enumerate}

\begin{table}[h]
	\fontsize{10}{10}\selectfont
    \caption{Adaptation Parameters for Spacecraft}
   \label{tab:adapt_param}
        \centering 
   \newcolumntype{R}{>{\raggedleft\arraybackslash}p{2cm}}
   \begin{tabular}{l | R | R } 
        Parameter & min & max \\
       \hline
      Spacecraft Initial Mass (kg) & 450 & 500 \\
      Probability of Actuator Failure ($p_{fail}$) & 0.5 & 0.5 \\
      Actuator Failure Range ($f_{min}$, $f_{max}$) & 0.50  &  1.0 \\
      Seeker Range Measurement Bias ($r_{bias}$) &  -0.05 & 0.05 \\
      Seeker Angle Measurement Bias ($a_{bias})$ & -0.05 & 0.05 \\
      Seeker Gaussian Angle Noise SD ($a_{\sigma}$) (mrad) & 1.0 & 1.0 \\
      Attitude Measurement Bias ($q_{bias}$) & 0.05 & 0.05 \\
      Angular Velocity Measurement Bias ($\omega_{bias})$ & 0.05 & 0.05 \\
      Center of Mass Variation (cm) & -10 & 10 \\
   \end{tabular}
\end{table}

\subsection{Equations of Motion}

The force $\mathbf{F}_{B}$ and torque $\mathbf{L}_{B}$ in the lander's body frame for a given commanded thrust depends on the placement of the thrusters in the lander structure. We can describe the placement of each thruster through a body-frame direction vector $\mathbf{d}$ and position vector $\mathbf{r}$, both in $\mathbb{R}^3$. The direction vector is a unit vector giving the direction of the body frame force that results when the thruster is fired.  The position vector gives the body frame location with respect to the  center of mass,  where the force resulting from the thruster firing is applied for purposes of computing torque, and in general the center of mass varies with time as fuel is consumed. For a lander with $k$ thrusters, the body frame force and torque associated with one or more  thrusters firing is then as shown in Equations \eqref{eq:Thruster_modela} and \eqref{eq:Thruster_modelb}, where $T_{cmd_{i}}\in[T_{min},T_{max}]$ is the commanded thrust for thruster $i$, $T_{min}$ and $T_{max}$ are a thruster's minimum and maximum thrust, $\mathbf{d}^{(i)}$ the direction vector for thruster $i$, and $\mathbf{r}^{(i)}$ the position of thruster $i$. The total body frame force and torque are calculated by summing the individual forces and torques.

\begin{subequations}
\begin{align}
	{\mathbf{F}_{B}}&={\sum_{i=1}^{k}\mathbf{d}^{(i)} T_{cmd}^{(i)}}\label{eq:Thruster_modela}\\
	{\mathbf{L}_{B}}&={\sum_{i=1}^{k}(\mathbf{r}^{(i)}-\mathbf{r}_\mathrm{com})\times\mathbf{F}_{B}^{(i)}}\label{eq:Thruster_modelb}
\end{align}
\end{subequations}

The dynamics model uses the lander's current attitude $\mathbf{q}$ to convert the body frame thrust vector to the inertial frame as shown in in Equation \eqref{eq:BtoN} where $[\mathbf{BN}](\mathbf{q})$ is the direction cosine matrix mapping the inertial frame to body frame obtained from the current attitude parameter $\mathbf{q}$.

\begin{equation}
	\label{eq:BtoN}
	\mathbf{F}_{N}=\left[\left[\mathbf{BN}\right](\mathbf{q})\right]^{T}\mathbf{F}_{B}
\end{equation}

The rotational velocities $\bm{\omega}_{B}$ are then obtained by integrating the Euler rotational equations of motion, as shown in Equation \eqref{eq:EulerRot}, where $\mathbf{L}_{B}$ is the body frame torque as given in Equation \eqref{eq:Thruster_modelb}, $\mathbf{L}_{env}$ is the body frame torque from external disturbances, and $\mathbf{J}$ is the lander's inertia tensor. Note we have included a term that models a rotation induced by a changing inertia tensor.

\begin{equation}
	\label{eq:EulerRot}
	\mathbf{J}{\dot{\bm{\omega}}_{B}}=-\Tilde{\bm{\omega}}_{B}\mathbf{J}\bm{\omega}_{B}-\mathbf{\dot{J}}\bm{\omega}+\mathbf{L}_{B}+\mathbf{L}_{B_{env}}
\end{equation}

The lander's attitude is then updated by integrating the differential kinematic equations shown in Equation \eqref{eq:diffeqom}, where the lander's attitude is parameterized using the quaternion representation and $\bm{\omega}_{i}$ denotes the $i^{th}$ component of the rotational velocity vector $\bm{\omega}_{B}$. 

\begin{equation}
    \label{eq:diffeqom}
    \begin{bmatrix} \dot{q_{0}} \\ \dot{q_{1}} \\ \dot{q_{2}} \\ \dot{q_{3}}\end{bmatrix} = \frac{1}{2}\begin{bmatrix} q_{0} & -q_{1} & -q_{2} & -q_{3}\\ q_{1} & q_{0} & -q_{3} & q_{2}\\ q_{2} & q_{3} & q_{0} & -q_{1} \\ q_{3} & -q_{2} & q_{1} & q_{0} \end{bmatrix} \begin{bmatrix} 0 \\ \omega_{0} \\ \omega_{1} \\ \omega_{2} \end{bmatrix}
\end{equation}

The translational motion is modeled as shown in \ref{eq:EQOMa} through \ref{eq:EQOMc}.

\begin{subequations}
\begin{align}
	{\Dot{\mathbf r}} &= {{\mathbf v}}\label{eq:EQOMa}\\
	{\Dot{\bf v}} &= \frac{{{\bf F}_{N}}}{m} + {{\bf a}_\text{env}} - g(\mathbf{r},a,b,c,\rho) +2\mathbf{\dot{r}}\times\bm{\omega}_a + (\bm{\omega}_a\times\mathbf{r})\times\bm{\omega}_a\label{eq:EQOMb}\\
	\Dot{m} &= -\frac{\sum_{i}^{k}\lVert{{\bf F}_{B}}^{(i)}\rVert}{I_\text{sp}g_\text{ref}} \label{eq:EQOMc}
\end{align}
\end{subequations}
Here  ${{\bf F}_{N}}^{(i)}$ is the inertial frame force as given in Eq.~\eqref{eq:BtoN}, $k$ is the number of thrusters, $g_\text{ref}=9.8$ $\text{m}/\text{s}^{2}$,  $\mathbf{r}$ is the spacecraft's position in the asteroid centered reference frame,  $g(\mathbf{r}_{a},a,b,c,\rho)$ is an ellipsoid gravity model as described in Reference(\citenum{scheeres2016orbital}), $\rho$ is the asteroid's density, $a,b,c$ are the asteroid's semi-axis lengths in meters, $I_\text{sp}=225$ s, and the spacecraft's mass is $m$.  ${\bf a}_\text{env}$ is a vector  representing solar radiation pressure. $\bm{\omega}_a$ is the asteroid's rotational velocity vector, which we compute as shown in Equations \eqref{eq:wa_a} through \eqref{eq:wa_f}, which uses the simplifying assumption that $J_x=J_y$ \cite{sawai2002control}. Here $\omega_o$ is the asteroid's spin rate and $\theta$ the nutation angle between the asteroid's spin axis and z-axis. We modified the equations from Reference (\citenum{sawai2002control}) to add the phase term $\phi$ to handle the case where the spacecraft starts the maneuver at an arbitrary point in the asteroid's rotational cycle. 

\begin{subequations}
\begin{align}
    \omega_{a_x} &= \omega_o\sin{\theta}\cos{(\omega_nt+\phi)}\label{eq:wa_a}\\
    \omega_{a_y} &= \omega_o\sin{\theta}\sin{(\omega_nt+\phi)}\label{eq:wa_b}\\
    \omega_{a_z} &= \omega_o\cos{\theta}\label{eq:wa_c}\\
    \omega_n &= \sigma\omega_o\cos{\theta}\label{eq:wa_d}\\
    {\bf J}&=\frac{m}{5}\begin{bmatrix} b^2+c^2 & 0 & 0 \\ 0 & a^2+c^2 & 0 \\ 0 & 0 & a^2+b^2\end{bmatrix}\\
    \sigma &= \frac{(J_z-J_x)}{J_x}\label{eq:wa_f}
\end{align}
\end{subequations}

The navigation system provides updates to the guidance system every 6 s, and we integrate the equations of motion using fourth order Runge-Kutta integration with a time step of 2 s.

\subsection{Guidance Law Development}

\subsection{RL Overview}

In the RL framework, an agent learns through episodic interaction with an environment how to successfully complete a task by learning a policy  that maps observations to actions. The environment initializes an episode by randomly generating a ground truth state, mapping this state to an observation, and passing the observation to the agent. These observations could be a corrupted version of the ground truth state (to model sensor noise) or could be raw sensor outputs such as Doppler radar altimeter readings, a multi-channel pixel map from an electro-optical sensor, or in our case, seeker angles and range to target.  The agent uses this observation to generate an action that is sent to the environment; the environment then uses the action and the current ground truth state to generate the next state and a scalar reward signal.  The reward and the observation corresponding to the next state are then passed to the agent. The process repeats until the environment terminates the episode, with the termination signaled to the agent via a done signal. Possible termination conditions include the agent completing the task, satisfying some condition on the ground truth state (such as altitude falling below zero), or violating a constraint.  
 
A Markov Decision Process (MDP) is an abstraction of the environment, which in a continuous state and action space, can be represented by a state space $\mathcal{S}$, an action space $\mathcal{A}$, a state transition distribution $\mathcal{P}(\mathbf{x}_{t+1}|\mathbf{x}_t,\mathbf{u}_t)$, and a reward function $r=\mathcal{R}(\mathbf{x}_t,\mathbf{u}_t))$, where $\mathbf{x} \in \mathcal{S}$ and $\mathbf{u} \in \mathcal{A}$, and $r$ is a scalar reward signal. We can also define a partially observable MDP (POMDP), where the state $\mathbf{x}$ becomes a hidden state, generating an observation $\mathbf{o}$ using an observation function $\mathcal{O}(\mathbf{x})$ that maps states to observations. The POMDP formulation is useful when the observation consists of raw sensor outputs, as is the case in this work.  In the following, we will refer to both fully observable and partially observable environments as POMDPs, as an MDP can be considered a POMDP with an identity function mapping states to observations.

The agent operates within an  environment defined by the POMDP, generating some action $\mathbf{u}_t$ based off of the observation $\mathbf{o}_t$, and receiving reward $r_{t+1}$ and next observation $\mathbf{o}_{t+1}$. Optimization involves maximizing the sum of (potentially discounted) rewards over the trajectories induced by the interaction between the agent and environment. Constraints such as minimum and maximum thrust, glide slope, attitude compatible with sensor field of view,  maximum rotational velocity, and terrain feature avoidance (such as targeting the bottom of a deep crater) can be included in the reward function, and will be accounted for when the policy is optimized. Note that there is no guarantee on the optimality of trajectories induced by the policy, although in practice it is possible to get close to optimal performance by tuning the reward function.

Reinforcement meta-learning differs from generic reinforcement learning in that the agent learns to quickly adapt to novel POMPDs by learning over a wide range of POMDPs. These POMDPs can include different environmental dynamics, actuator failure scenarios, mass and inertia tensor variation, and varying amounts of sensor distortion. Learning within the RL meta-learning framework results in an agent that can quickly adapt to novel POMDPs, often with just a few steps of interaction with the environment. There are multiple approaches to implementing meta-RL.  In Reference (\citenum{finn2017model}), the authors design the objective function to explicitly make the model parameters transfer well to new tasks. In Reference (\citenum{mishra2018simple}), the authors demonstrate state of the art performance using temporal convolutions with soft attention. And  in Reference (\citenum{frans2017meta}), the authors use a hierarchy of policies to achieve meta-RL. In this proposal, we use a different approach\cite{wang2016learning} using a recurrent policy and value function. Note that it is possible to train over a wide range of POMDPs using a non-meta RL algorithm\cite{gaudet2018deep}. Although such an approach typically results in a robust policy, the policy cannot adapt in real time to novel environments. 

In this work, we  implement metal-RL using proximal policy optimization (PPO) \cite{schulman2017proximal} with both the policy and value function implementing recurrent layers in their networks.  To understand how recurrent layers result in an adaptive agent, consider that given some ground truth agent position, velocity, attitude, and rotational velocity $\mathbf{x}_{t}$, and action vector $\mathbf{u}_{t}$ output by the agent's policy, the next state $\mathbf{x}_{t+1}$ and observation $\mathbf{o}_{t+1}$ depends not only on $\mathbf{x}_{t}$ and $\mathbf{u}_{t}$, but also on the ground truth agent mass, inertia tensor, and external forces acting on the agent. Consequently, during training, the hidden state of a network's recurrent network evolves differently depending on the observed sequence of observations from the environment and actions output by the policy. Specifically, the trained policy's hidden state captures unobserved (potentially time-varying) information such as external forces that are useful in minimizing the cost function. In contrast, a non-recurrent policy (which we will refer to as an MLP policy), which does not maintain a persistent hidden state vector, can only optimize using a set of current observations, actions, and advantages, and will tend to under-perform a recurrent policy on tasks with randomized dynamics, although as we have shown in (Reference \citenum{gaudet2018deep}), training with parameter uncertainty can give good results using an MLP policy, provided the parameter uncertainty is not too extreme.  After training, although the recurrent policy's network weights are frozen, the hidden state will continue to evolve in response to a sequence of observations and actions, thus making the policy adaptive.  In contrast, an MLP policy's behavior is fixed by the network parameters at test time.

The PPO algorithm used in this work  is a  policy gradient algorithm which has demonstrated state-of-the-art performance for many RL benchmark problems. PPO approximates the TRPO optimization process\cite{schulman2015trust} by accounting for the policy adjustment constraint with a clipped objective function. The objective function used with PPO can be expressed in terms of the probability ratio $p_{k}({\bm \theta})$ given by Eq.~\eqref{eq:clipr}, where $\pi_\theta$ is the policy parameterized by parameter vector $\theta$.
\begin{equation}
\label{eq:clipr} 
p_{k}({\bm \theta})=\frac{\pi_{{\bm \theta}}({\bf u}_{k}|{\bf o}_{k})}{\pi_{{\bm \theta}_\text{old}}({\bf u}_{k}|{\bf o}_{k})}
\end{equation}
The PPO objective function is then given in Eq.~\eqref{eq:ppoloss}:
\begin{equation}
\label{eq:ppoloss}
J({\bm \theta})=\mathbb{E}_{p({\bm \tau})}\left[\mathrm{min}\left[p_{k}({\bm \theta}) , \mathrm{clip}(p_{k}({\bm \theta}) , 1-\epsilon, 1+\epsilon)\right]A^{\pi}_{\bf w}({\bf o}_{k},{\bf u}_{k})\right]
\end{equation}
This clipped objective function has been shown to maintain a bounded KL divergence with respect to the policy distributions between updates, which aids convergence by insuring that the policy does not change drastically between updates. Our implementation of PPO uses an approximation to the advantage function that is the difference between the empirical return and a state value function baseline, as shown in Equation \ref{eq:ppo_adv}:
\begin{equation}
\label{eq:ppo_adv}
    A^{\pi}_{\bf w}(\mathbf{x}_{k},\mathbf{u}_{k})=\left[\sum_{\ell=k}^{T}\gamma^{\ell-k}r(\bf o_{\ell},\bf u_{\ell})\right]-V_{\bf w}^{\pi}(\mathbf{x}_{k})
\end{equation}
Here the value function $V_{\bf w}^{\pi}$ parameterized by vector $\mathbf w$ is learned using the cost function given by Eq.~\eqref{eq:vf_ppo}, where $\gamma$ is a discount rate applied to rewards generated by reward function $\mathcal{R}(\mathbf{o},\mathbf{u})$.  The discounting of rewards improves optimization performance by improving temporal credit assignment.
\begin{equation}
\label{eq:vf_ppo}
L(\mathbf{w})=\sum_{i=1}^{M}\left(V_{\mathbf{w}}^{\pi}({\bf o}_k^i)-\left[\sum_{\ell=k}^{T}\gamma^{\ell-k}\mathcal{R}({\bf u}_{\ell}^i,{\bf o}_{\ell}^i)\right]\right)^2
\end{equation}
In practice, policy gradient algorithms update the policy using a batch of trajectories (roll-outs) collected by interaction with the environment. Each trajectory is associated with a single episode, with a sample from a trajectory collected at step $k$ consisting of observation ${\bf o}_{k}$, action ${\bf u}_{k}$, and reward $r_k=\mathcal{R}({\bf o}_k,{\bf u}_k)$. Finally, gradient ascent is performed on ${\bm \theta}$ and gradient decent on ${\bf w}$ and update equations are given by
\begin{align}\label{loss}
{\bf w}^+&={\bf w}^--\beta_{{\bf w}}\nabla_{{\bf w}} \left. L({\bf w})\right|_{{\bf w}={\bf w}^-}\\
{\bm \theta}^+&={\bm \theta}^-+\beta_{{\bm \theta}} \left. \nabla_{\bm \theta}J\left({\bm \theta}\right)\right|_{{\bm \theta}={\bm \theta}^-}
\end{align}
where $\beta_{{\bf w}}$ and $\beta_{{\bm \theta}}$ are the learning rates for the value function, $V_{\bf w}^{\pi}\left({\bf o}_k\right)$, and policy, $\pi_{\bm \theta}\left({\bf u}_k|{\bf o}_k\right)$, respectively.

In our implementation, we dynamically adjust the clipping parameter $\epsilon$ to target a KL divergence between policy updates of 0.001. The policy and value function are learned concurrently, as the estimated value of a state is policy dependent. We use a Gaussian distribution with mean $\pi_{\bm \theta}({\bf o}_{k})$ and a diagonal covariance matrix for the action distribution in the policy.  Because the log probabilities are calculated using the exploration variance, the degree of exploration automatically adapts during learning such that the objective function is maximized.

\subsection{Guidance Law Optimization}

\begin{figure}[h]
\begin{center}
\includegraphics[width=.9\linewidth]{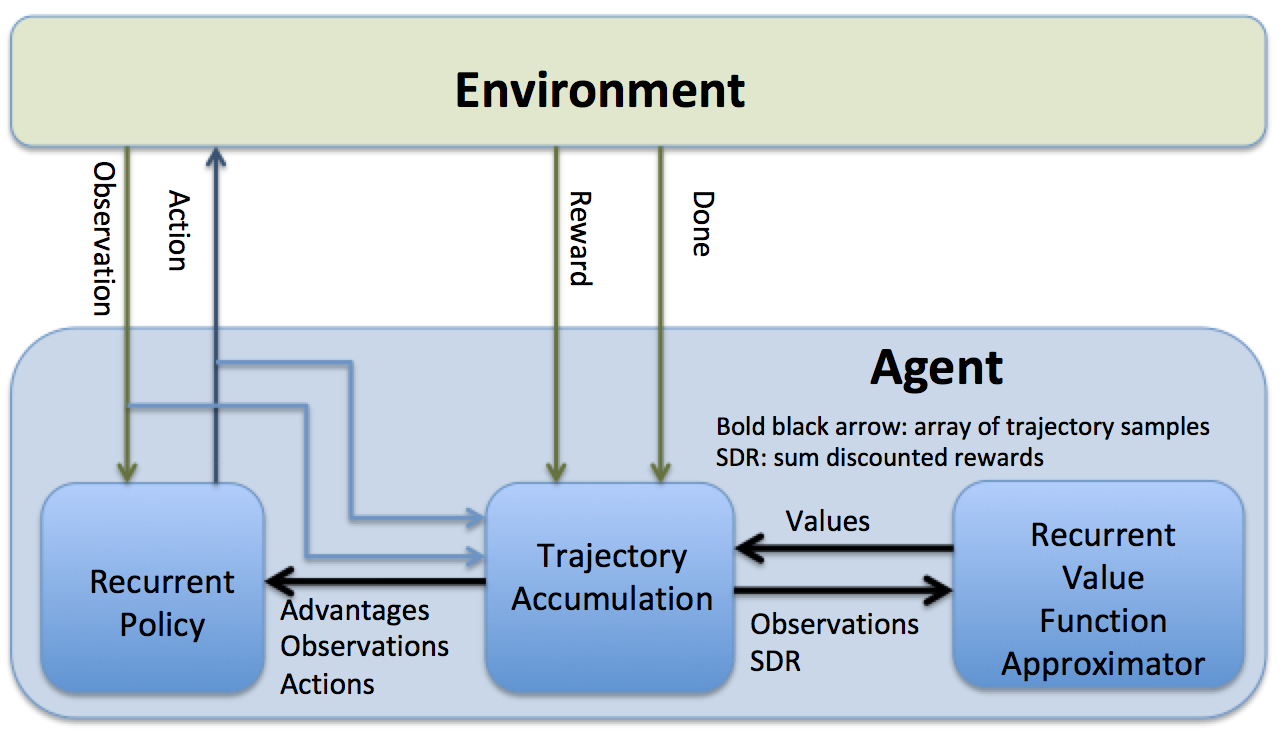}
\caption{Agent-Environment Interface}
\label{fig:a_e_i}
\end{center}
\end{figure}

A simplified view of the agent and environment are shown in Figure \ref{fig:a_e_i}. The environment instantiates the system dynamics model, reward function, spacecraft model, and thruster model. The observation given to the agent is shown in Eq.~\eqref{eq:obs}. Referring to Eqs.~\ref{eq:seeker3} and \ref{eq:seeker4}, the agent's observation include the seeker angles $\theta_{u}$ and $\theta_{v}$ and their time derivatives $\dot\theta_{u}$ and $\dot\theta_{v}$. The observation further includes the spacecraft's rotational velocity $\boldsymbol{\omega}$, the velocity magnitude tracking error $v_\text{error}$, and the change in attitude since the start of the maneuver $\boldsymbol{dq}$.  The velocity tracking error is given in Eqs.~\ref{eq:vref} through \ref{eq:verr}, where $r$ is the measured range to target, and $v_c$ the closing velocity, measured using either the Doppler effect (if Doppler LIDAR is used) or by filtering $r$. The use of the change in attitude rather than absolute attitude  makes the observation landing site invariant in that we can optimize using any target location on the asteroid's surface and the policy will generalize to any other target location.

\begin{equation}
    \label{eq:obs}
    \mathrm{obs} = \begin{bmatrix} \theta_u & \theta_v & \dot\theta_u & \dot\theta_v &  v_{\text{error}} & t_{\text{go}}  & \boldsymbol{dq} & \boldsymbol{\omega} &\end{bmatrix} 
\end{equation}

\begin{subequations}
\begin{align}
	v_{\text{ref}}&=v_{o}(1-\text{exp}(-\frac{t_{\text go}}{\tau})\label{eq:vref}\\
	t_{\text{go}}&=\frac{r}{v_{c}}\label{eq:tgo}\\
	{v}_\text{error}&=v_{\text{ref}}-v_c\label{eq:verr}
\end{align}
\end{subequations}

The action space is in $\mathbb{R}^{k}$, where k is the number of thrusters.  The discretized agent action $\mathbf{u} \in [0,1]$ is used to index Table \ref{tab:thrusters}, where if the discretized action is 1, it is used to compute the body frame force and torque contributed by that thruster. The discretization is necessary since the policy uses a Gaussian action distributions.  Policy actions greater than zero are mapped to one (fire thruster), and actions less than zero are mapped to zero (don't fire).

The policy and value functions are implemented using four layer neural networks with tanh activations on each hidden layer. Layer 2 for the policy and value function is a recurrent layer implemented as a gated recurrent unit \cite{chung2015gated}. The network architectures are as shown in Table \ref{tab:NN}, where $n_{\mathrm{hi}}$ is the number of units in layer $i$, $\mathrm{obs\_dim}$ is the observation dimension, and $\mathrm{act\_dim}$ is the action dimension. The policy and value functions are periodically updated during optimization after accumulating trajectory rollouts of 30 simulated episodes.

\begin{table}[h]
	\fontsize{10}{10}\selectfont
    \caption{Policy and Value Function network architecture}
   \label{tab:NN}
        \centering 
   \newcolumntype{R}{>{\raggedleft\arraybackslash}p{1.8cm}}
   \begin{tabular}{l | R | c | R | c } 
      \hline 
       & \multicolumn{2}{c}{Policy Network}\vline & \multicolumn{2}{c}{Value Network}\\
       \hline
       Layer & \# units & activation & \# units & activation \\
       \hline
      hidden 1      & $10 * \mathrm{obs\_dim}$ & tanh & $10 * \mathrm{obs\_dim}$ & tanh \\
      hidden 2      & $\sqrt{n_{\mathrm{h1}} * n_{\mathrm{h3}}}$ & tanh & $\sqrt{n_{\mathrm{h1}} * n_{\mathrm{h3}}}$ & tanh\\
      hidden 3      & $10 * \mathrm{act\_dim}$ & tanh & 5 & tanh \\
      output        & $\mathrm{act\_dim}$ & linear & 1 & linear \\
      \hline
   \end{tabular}
\end{table}

Since it is unlikely that the agent will achieve a good landing by random exploration, we provide shaping rewards\cite{ng2003shaping} for nullifying the seeker angles and tracking the reference velocity magnitude, with the seeker angle tracking error shown in Equation (\ref{eq:serr}).

\begin{subequations}
\begin{align}
	{s}_{\text{error}}&=\|[\theta_u,\theta_v]\|\label{eq:serr}
\end{align}
\end{subequations}

Finally, we provide a terminal reward bonus when the spacecraft executes a good landing (see below). The reward function is then given by Equation \eqref{eq:reward_func}, where the various terms are described in the following:

\begin{enumerate}
    \item $\alpha$ weights a term penalizing the error in tracking the target velocity.
    \item $\beta$ weights a term penalizing non-zero seeker angles, i.e. $s_\text{error}$.
    \item $\gamma$ weights a term penalizing control effort.
    \item $\eta$ is a constant positive term that encourages the agent to keep making progress along the trajectory.
    \item $\zeta$ is a bonus given for a successful landing, where the spacecraft's terminal position and velocity are all within specified limits.  The limits are $\|\mathbf{r}\|=1$ $\text{m}$, $\|\mathbf{v}\|=0.1$, $\text{m/s}$, and all components of angular velocity less than 0.025 rad/sec
    \item $\kappa$ is a penalty for exceeding any constraint. We impose a rotational velocity constraint of 0.10 rad/sec for all three rotational axes. We also constrain the target to remain in the seeker's field of regard.  If a constraint is violated, the episode terminates.
\end{enumerate}

\begin{equation}
 \begin{aligned}
 \label{eq:reward_func}
r &= \alpha v_{\text error}+ \beta s_{\text error} + \gamma\|{\bf T}\|+ \eta +\zeta(\text{good landing})+\kappa(\text{constraint violation})
\end{aligned}
\end{equation}

Initial hyperparameter settings are shown in Table \ref{tab:HPS}.

\begin{table}[h]
	\fontsize{10}{10}\selectfont
    \caption{Hyperparameter Settings}
   \label{tab:HPS}
        \centering 
   \begin{tabular}{ c | c | c | c | c | c | c | c  } 
      \hline
      $v_{o}$ (m/s) & $\tau$ (s)  & $\alpha$  & $\beta$   &  $\gamma$ & $\eta$ & $\zeta$ & $\kappa$\\
      \hline
       0.5 & 300  & -0.5  & -0.5  &  -0.05 & 0.01 & 10 & -50\\
      
   \end{tabular}
\end{table}

\section{Experiments}

\subsection{Optimization Results}

\begin{figure}[h]
\begin{center}
\includegraphics[width=.9\linewidth]{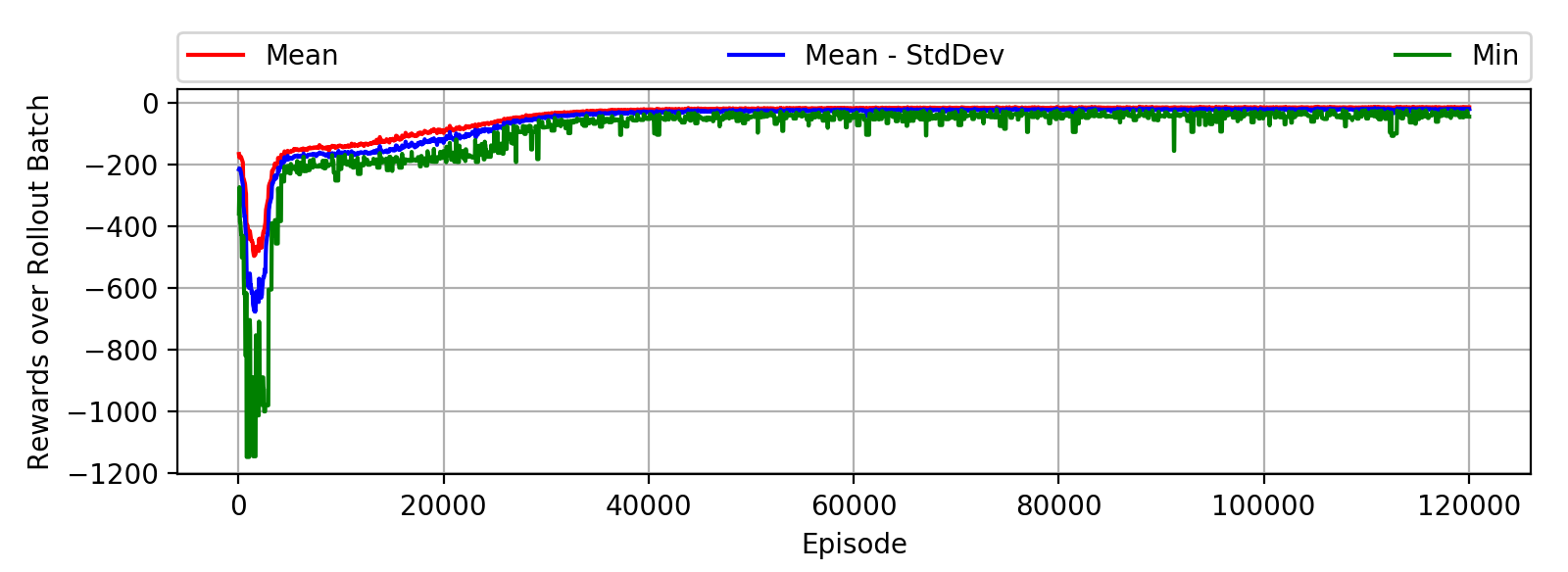}
\caption{Optimization Rewards Learning Curves}
\label{fig:rewards}
\end{center}
\end{figure}

\begin{figure}[h]
\begin{center}
\includegraphics[width=.9\linewidth]{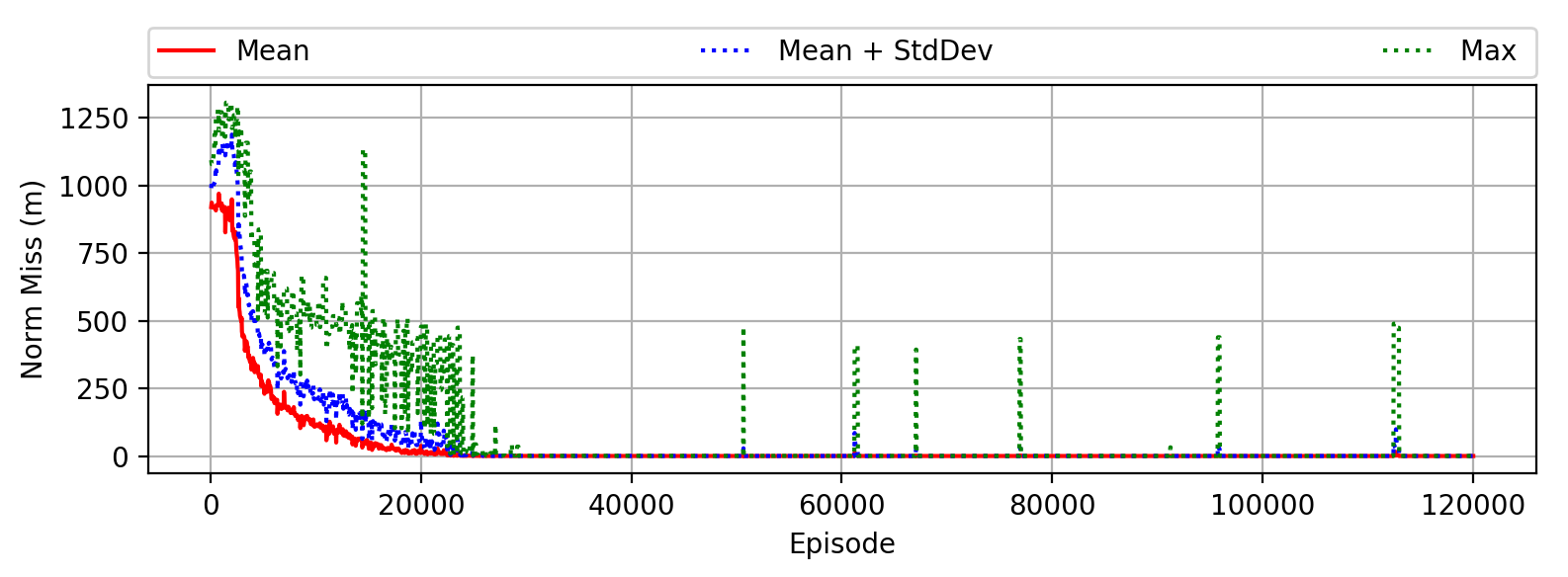}
\caption{Optimization Terminal Position Learning Curves}
\label{fig:rf}
\end{center}
\end{figure}

\begin{figure}[h!]
\begin{center}
\includegraphics[width=.9\linewidth]{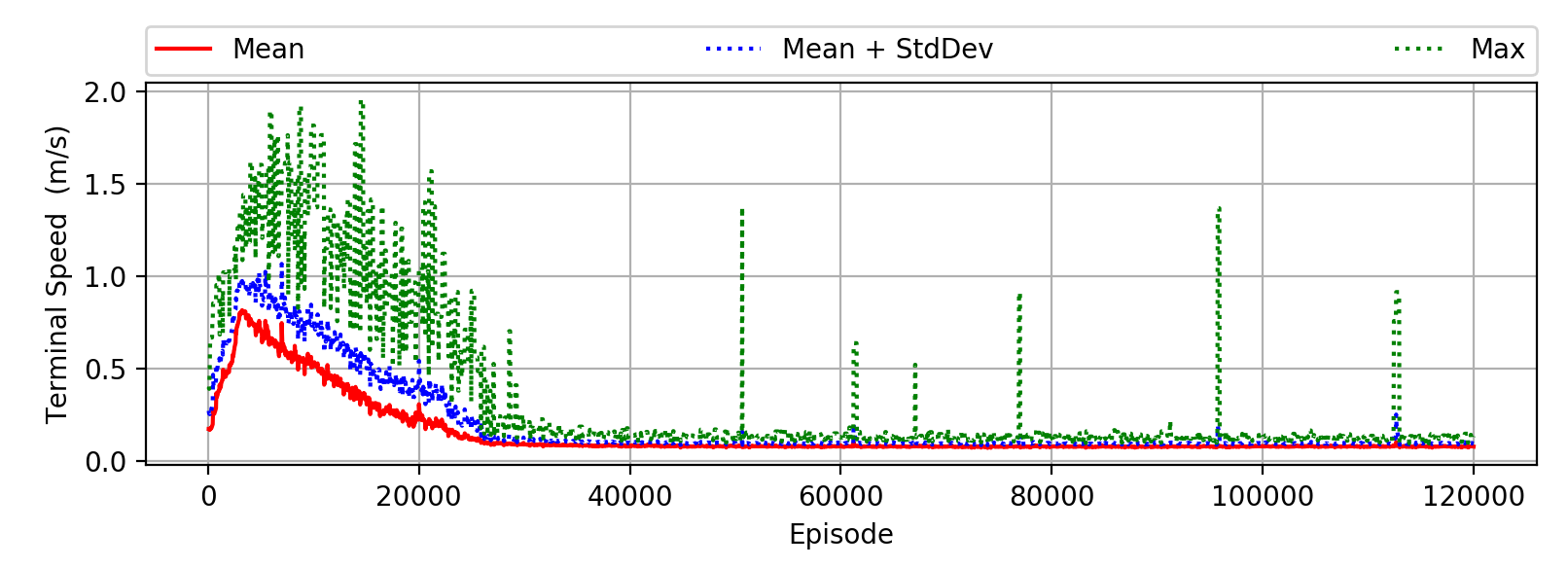}
\caption{Optimization Terminal Velocity Learning Curves}
\label{fig:vf}
\end{center}
\end{figure}

We optimize the policy using the initial condition parameters given in Table \ref{tab:initial_conditions} and system uncertainty given in Tables \ref{tab:ast_param} and \ref{tab:adapt_param}, with the exception that the optimization uses a simplified asteroid dynamics model where the asteroid is modeled as a sphere for purposes of calculating the gravitational field, with mass from $1\times10^{-10}$ to $150\times10^{-10}$.  The simplified model significantly reduces optimization time. However, for policy testing, we use the ellipsoid gravity model. Optimization curves are shown in Figures \ref{fig:rewards} through \ref{fig:vf}. Fig.~\ref{fig:rewards}  plots reward statistics over each rollout batch of 30 episodes, Fig.~\ref{fig:rf} plots the terminal miss distance, and Fig.~\ref{fig:vf} plots the norm of the lander's terminal velocity. Note that during optimization there are rare catastrophic failures identifiable by the spikes in Figures~\ref{fig:rf} and \ref{fig:vf}.  These are due to the fact that the exploration standard deviation was still around 0.20 (20\% of maximum thrust) and slowly dropping when we terminated the optimization at 120 episodes. This exploration can occasionally result in the policy taking an action that results in catastrophic failure, which is why we  turn exploration off for the deployed policy.

\subsection{Testing Results}

\begin{figure}[h!]
\begin{center}
\includegraphics[width=.8\linewidth]{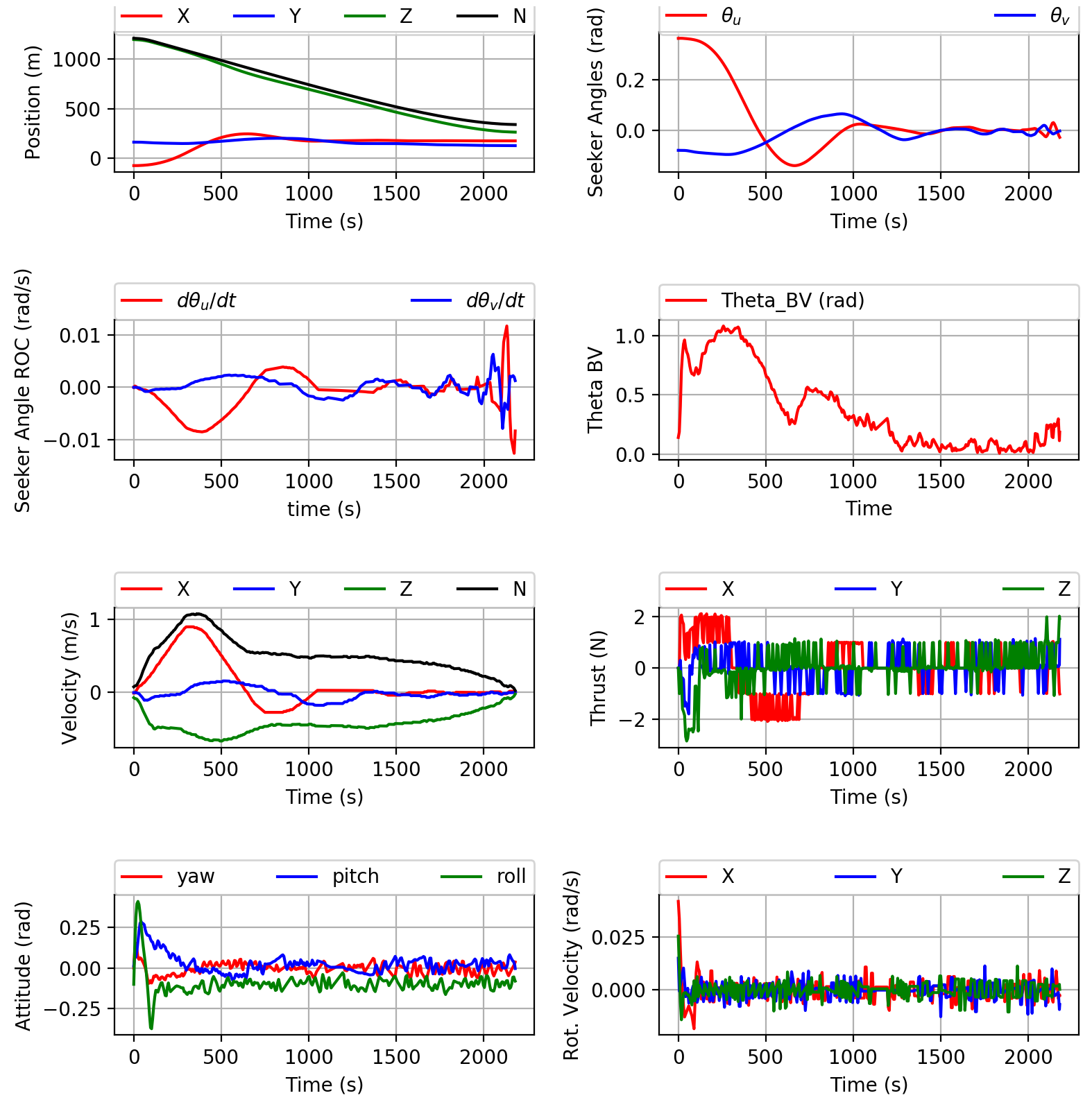}
\caption{Sample trajectory}
\label{fig:traj}
\end{center}
\end{figure}

For testing, we use the initial conditions given in Table \ref{tab:initial_conditions}, the high fidelity asteroid model (ellipsoid gravity field) with parameters as given in Table \ref{tab:ast_param}, and spacecraft parameters given in Table \ref{tab:adapt_param}.  To be clear, each episode has some amount of sensor noise, probability of actuator failure, and initial center of mass variation.  For each episode, the landing site is randomly chosen on the ellipsoid.  Note that optimization directly targeted the landing site, during test we target a location 10m from the landing site, which is our desired final location.  This is accomplished by adjusting the range measurement such that an adjusted range of zero corresponds to a distance $d$ from the landing site. We ran tests for $d=0\text{m}$ to $d=10\text{m}$, and the results in the following were generated with $d=10\text{m}$. Test results are given in Table \ref{tab:results}, which are computed from 5000 simulated episodes. Note that the rotational velocity row in Table \ref{tab:results} gives the rotational velocity vector element with the worst performance. The "Good Landing" row gives the percentage of episodes where the terminal miss distance was less than 1m, terminal speed less than 10cm/s, and all elements of the terminal rotational velocity less than 0.025 rad/s. Although we did not achieve good landings 100\% of the time, from the maximum values in Table \ref{tab:results}, we see that the constraints were only slightly violated.   A sample trajectory is shown in Figure \ref{fig:traj}, where the position subplot is in the asteroid centered reference frame. The "Theta\_BV" plot shows the angle (in radians) between the spacecraft's Z-axis and the spacecraft's velocity vector. Note that when the environmental dynamics are such that maximum thrust is not required, the policy fires only a single thruster on a given side of the spacecraft, resulting in a 1N thrust. This saves fuel at the expense of inducing rotation, which is compensated for by firing the opposing thruster on the opposite side at some future time.

\begin{table}[H]
	\fontsize{10}{10}\selectfont
    \caption{Performance}
   \label{tab:results}
        \centering 
   \newcolumntype{R}{>{\raggedleft\arraybackslash}p{3cm}}
   \begin{tabular}{ l | R | R | R   } 
      \hline
      Parameter & Mean & Std & Max \\
      \hline
      Terminal Position (cm) & 2.7 & 2.5 & 20.2  \\
      Terminal Velocity (cm/s) & 3.1 & 1.5 & 11.0   \\
      Rotational Velocity (mrad/s) & 0.0 & 3.6 & 25.3 \\
      Good Landing (\%) & 99.78 & N/A & N/A \\
      Fuel (kg) & 1.43 & 0.25 & 2.80 \\
   \end{tabular}
\end{table}

\subsection{Increased Parameter Variation}
We ran additional simulations where we increased the parameter variation limits in Table \ref{tab:adapt_param}, as shown below, forcing the guidance system to adapt to a wider range of conditions. Although performance was slightly impacted, the results were close to that given in Table \ref{tab:results}. The worst performance achieved landing constraints 99.5\% of the time, and constraint violations were minimal.

\begin{enumerate}
\item\underline{Sensor Noise} We increased sensor $r_{bias}$ and $a_{bias}$ from 0.05 to 0.1. Attitude and rotational velocity sensor bias were both increased from 5\% to 10\%.  

\item\underline{Reduced Actuator Failure Lower Limit} The lower limit of actuator failure range $f_{min}$ is decreased to 0.40. At $f_{min}=0.40$, the maximum recorded force acting on the spacecraft was 82\% of the spacecraft's maximum thrust capability. Further decreasing $f_{min}$ resulted in inadequate thrust to effectively counter the Coriolis and centrifugal forces, resulting in constraint violation and early episode termination. 

\item\underline{Spin Rate} The asteroid's spin rate is increased to $0.6\times10^{-4} \text{rad/s}$, which increased the maximum recorded force acting on the spacecraft to 1.4N, 93\% of the spacecraft's maximum thrust capability taking into account actuator failure (1.5N). Further increases in the spin rate led to occasional constraint violation and early episode termination.

\item\underline{Center of Mass Variation} The spacecraft's initial center of mass variation is increased from 0.1m (10\%) to 0.2m (20\%).

\end{enumerate}

\subsection{Small Lander Experiment}

We re-ran optimization and testing using a miniaturized spacecraft configuration that has 8 times the volume of a standard cubesat to give sufficient volume for inclusion of sample collection hardware. This would be deployed by a hovering spacecraft to travel to the asteroid's surface for sample collection while the sample collection site is illuminated by a targeting laser from the hovering spacecraft, as described in the Concept of Operations section. The spacecraft is modeled as a cube with 20cm sides, a nominal mass of 5kg, and each thruster providing 10 mN of thrust. Thruster locations were modified  and spacecraft adaptation parameters were adjusted to take into account the new moments of inertia.  Performance was similar to that of the experiments described in the preceding, and we expect the approach could be scaled to a standard cubesat that could be used to implement surface navigation beacons.

\subsection{Discussion}

Overall performance met our goals, although fuel consumption was on the high side. We were able to improve fuel efficiency by a factor of two by using a more complex thruster configuration with dedicated attitude control thrusters, but this was less robust to actuator failure. Fuel efficiency could be significantly improved by putting the spacecraft on an initial heading that puts it on a collision triangle with the target landing site, however this would require the guidance system to be able to predict the target site's future location in the inertial asteroid centered reference frame. Specifically, instead of "chasing" the target, the spacecraft could be put on a collision triangle with the target, and the trajectory would terminate at the future target location.  This capability would also allow landings with much higher asteroid spin rates without increasing the spacecraft's thrust capability, and robustness to actuator failure would be increased.   We leave this to future work.

\section{Implementation Considerations}

In this work we considered an ideal seeker that perfectly tracked the target from a stabilized (inertial) platform.  When this guidance system is implemented on a cubesat, miniaturization of the seeker hardware is critical. First, note that since the GNC system only requires changes in attitude that have accumulated from the start of a maneuver, rather than use a star tracker, we can measure the difference between a gyroscope stabilized reference frame and the spacecraft body frame. This should aid with miniaturization.  Further, the equipment required to measure this change in attitude as well as the rotational velocity is currently implemented in smart phones, so the space and weight requirements are minimal.  It would then appear that the primary challenge would be miniaturization of the seeker actuators that change the two seeker angles and the seeker platform stabilization actuators. 

Considerable hardware simplfication and miniaturization would be possible using a strap down seeker implementation, where both the seeker platform and the seeker's boresight axis are fixed in the body frame. In a strap down seeker, the platform is stabilized by measuring the change in the spacecraft's attitude during a maneuver, and applying the appropriate rotation operator to the measured seeker angles.  And rather than mechanically adjusting the seeker boresight axis direction, the boresight is fixed in the body frame and the seeker angles are calculated based off of the target's position in the seeker's field of regard. However, due to the requirement of having the LIDAR altimeter aligned with the seeker's optical axis, the seeker platform must be mechanically stabilized and the seeker itself must be mechanically pointed so that the seeker camera's optical axis and the LIDAR rangefinder both point along the same direction vector, with the range finder giving the range to the location in the center of the camera's field of view. This precludes a strap down seeker implementation, where both the seeker platform and the seeker's boresight axis are fixed in the body frame.

In theory, it should be possible to eliminate the LIDAR range finder hardware by employing two cameras, i.e., giving the seeker binocular vision, where the location of the target in the two camera's fields of view will differ, with the difference depending on range. This would allow the implementation of a full strap down seeker with no moving parts. Moreover, note that the simulation results in the Experiments section show that there is no need for high fidelity sensors as the guidance law can tolerate considerable sensor distortion.  Note that we have run some preliminary experiments that show good seeker performance with significant camera slant, radial distortion, and tangential  distortion, so the use of a small camera should not be an issue. Finally, note that there are existing space certified propulsion systems designed to be implemented on a cubesat that would provide sufficient thrust for the landing scenarios described in this work.

\section{Conclusion}

We formulated a particularly difficult problem: precision maneuvers around an asteroid with unknown dynamics, starting from a large range of initial condition uncertainty, accounting for actuator failure, center of mass variation, and sensor noise, and using raw sensor measurements. We created a high fidelity 6-DOF simulator that synthesized asteroid models with randomized parameters. where the asteroid is modeled as a uniform density ellipsoid that in general is not rotating about a principal axis, resulting in time varying dynamics. The problem was solved by optimizing an adaptive policy that mapped sensor measurements directly to actuator commands.  The policy was optimized using reinforcement meta-learning, where the policy and value function each contained a recurrent hidden layer. Different asteroid characteristics, center of mass variation, actuator failure and sensor noise represented different POMDPs during optimization, and maximizing discounted rewards required the policy recurrent layer's state to evolve based off of a history of observations and actions, effectively adapting to the environment. The optimized policy was extensively tested, with satisfactory results.  Finally, we suggested a concept of operations where the spacecraft deploys multiple landers to act as beacons and collect samples, while the spacecraft remains safely in a hovering position where it can orchestrate the lander's maneuvers using a targeting laser. One area of future work will focus on the seeker technology, where we will explore locking onto terrain features rather than beacons, and replacing the LIDAR rangefinder with binocular vision. Another focus of future work will be the ability to put the spacecraft on a collision triangle with the target, which will improve both fuel efficiency and robustness to rapid asteroid spin rates.   The approach discussed in this work would also be applicable to orbital rendezvous and Lunar landing, particularly Lunar sortie missions where the landing site is already marked with beacons.

\bibliographystyle{AAS_publication}   
\bibliography{references}   

\begin{thebibliography}{10}

\bibitem{gaskell2008characterizing}
R.~Gaskell, O.~Barnouin-Jha, D.~J. Scheeres, A.~Konopliv, T.~Mukai, S.~Abe,
  J.~Saito, M.~Ishiguro, T.~Kubota, T.~Hashimoto, {\em et~al.},
  ``Characterizing and navigating small bodies with imaging data,''  {\em
  Meteoritics \& Planetary Science}, Vol.~43, No.~6, 2008, pp.~1049--1061.

\bibitem{udrea2012sensitivity}
B.~Udrea, P.~Patel, and P.~Anderson, ``Sensitivity Analysis of the Touchdown
  Footprint at (101955) 1999 RQ36,''  {\em Proceedings of the 22nd AAS/AIAA
  Spaceflight Mechanics Conference}, Vol.~143, 2012.

\bibitem{schutz2004statistical}
B.~Schutz, B.~Tapley, and G.~H. Born, {\em Statistical orbit determination}.
\newblock Elsevier, 2004.

\bibitem{d1997optimal}
C.~D'Souza and C.~D'Souza, ``An optimal guidance law for planetary landing,''
  {\em Guidance, Navigation, and Control Conference}, 1997, p.~3709.

\bibitem{gaudet2019adaptive}
B.~Gaudet and R.~Linares, ``Adaptive Guidance with Reinforcement
  Meta-Learning,''  {\em arXiv preprint arXiv:1901.04473}, 2019.

\bibitem{prabhakar2018trajectory}
N.~Prabhakar, A.~Painter, R.~Prazenica, and M.~Balas, ``Trajectory-Driven
  Adaptive Control of Autonomous Unmanned Aerial Vehicles with Disturbance
  Accommodation,''  {\em Journal of Guidance, Control, and Dynamics}, Vol.~41,
  No.~9, 2018, pp.~1976--1989.

\bibitem{huang2019mars}
Y.~Huang, S.~Li, and J.~Sun, ``Mars entry fault-tolerant control via neural
  network and structure adaptive model inversion,''  {\em Advances in Space
  Research}, Vol.~63, No.~1, 2019, pp.~557--571.

\bibitem{han2015adaptive}
Y.~Han, J.~D. Biggs, and N.~Cui, ``Adaptive fault-tolerant control of
  spacecraft attitude dynamics with actuator failures,''  {\em Journal of
  Guidance, Control, and Dynamics}, Vol.~38, No.~10, 2015, pp.~2033--2042.

\bibitem{furfaro2018deep}
R.~Furfaro, I.~Bloise, M.~Orlandelli, P.~Di~Lizia, F.~Topputo, and R.~Linares,
  ``Deep Learning for Autonomous Lunar Landing,''  {\em 2018 AAS/AIAA
  Astrodynamics Specialist Conference}, 2018, pp.~1--22.

\bibitem{furfaro2018recurrent}
R.~Furfaro, I.~Bloise, M.~Orlandelli, P.~Di~Lizia, F.~Topputo, and R.~Linares,
  ``A Recurrent Deep Architecture for Quasi-Optimal Feedback Guidance in
  Planetary Landing,''  {\em IAA SciTech Forum on Space Flight Mechanics and
  Space Structures and Materials}, 2018, pp.~1--24.

\bibitem{furfaro2017waypoint3}
R.~Furfaro and R.~Linares, ``Waypoint-based generalized ZEM/ZEV feedback
  guidance for planetary landing via a reinforcement learning approach,''  {\em
  3rd International Academy of Astronautics Conference on Dynamics and Control
  of Space Systems, DyCoSS 2017}, Univelt Inc., 2017, pp.~401--416.

\bibitem{gaudet2018deep}
B.~Gaudet, R.~Linares, and R.~Furfaro, ``Deep Reinforcement Learning for Six
  Degree-of-Freedom Planetary Powered Descent and Landing,''  {\em arXiv
  preprint arXiv:1810.08719}, 2018.

\bibitem{siouris2004missile}
G.~M. Siouris, {\em Missile guidance and control systems}.
\newblock Springer Science \& Business Media, 2004.

\bibitem{2019arXiv190602113G}
B.~{Gaudet}, R.~{Furfaro}, and R.~{Linares}, ``{A Guidance Law for Terminal
  Phase Exo-Atmospheric Interception Against a Maneuvering Target using
  Angle-Only Measurements Optimized using Reinforcement Meta-Learning},''  {\em
  arXiv e-prints}, Jun 2019, p.~arXiv:1906.02113.

\bibitem{thrun2005probabilistic}
S.~Thrun, W.~Burgard, and D.~Fox, {\em Probabilistic robotics}.
\newblock MIT press, 2005.

\bibitem{scheeres2016orbital}
D.~J. Scheeres, {\em Orbital motion in strongly perturbed environments:
  applications to asteroid, comet and planetary satellite orbiters}.
\newblock Springer, 2016.

\bibitem{sawai2002control}
S.~Sawai, D.~Scheeres, and S.~Broschart, ``Control of hovering spacecraft using
  altimetry,''  {\em Journal of Guidance, Control, and Dynamics}, Vol.~25,
  No.~4, 2002, pp.~786--795.

\bibitem{finn2017model}
C.~Finn, P.~Abbeel, and S.~Levine, ``Model-agnostic meta-learning for fast
  adaptation of deep networks,''  {\em arXiv preprint arXiv:1703.03400}, 2017.

\bibitem{mishra2018simple}
N.~Mishra, M.~Rohaninejad, X.~Chen, and P.~Abbeel, ``A simple neural attentive
  meta-learner,''  2018.

\bibitem{frans2017meta}
K.~Frans, J.~Ho, X.~Chen, P.~Abbeel, and J.~Schulman, ``Meta learning shared
  hierarchies,''  {\em arXiv preprint arXiv:1710.09767}, 2017.

\bibitem{wang2016learning}
J.~X. Wang, Z.~Kurth-Nelson, D.~Tirumala, H.~Soyer, J.~Z. Leibo, R.~Munos,
  C.~Blundell, D.~Kumaran, and M.~Botvinick, ``Learning to reinforcement
  learn,''  {\em arXiv preprint arXiv:1611.05763}, 2016.

\bibitem{schulman2017proximal}
J.~Schulman, F.~Wolski, P.~Dhariwal, A.~Radford, and O.~Klimov, ``Proximal
  policy optimization algorithms,''  {\em arXiv preprint arXiv:1707.06347},
  2017.

\bibitem{schulman2015trust}
J.~Schulman, S.~Levine, P.~Abbeel, M.~Jordan, and P.~Moritz, ``Trust region
  policy optimization,''  {\em International Conference on Machine Learning},
  2015, pp.~1889--1897.

\bibitem{chung2015gated}
J.~Chung, C.~Gulcehre, K.~Cho, and Y.~Bengio, ``Gated feedback recurrent neural
  networks,''  {\em International Conference on Machine Learning}, 2015,
  pp.~2067--2075.

\bibitem{ng2003shaping}
A.~Y. Ng, {\em Shaping and policy search in reinforcement learning}.
\newblock PhD thesis, University of California, Berkeley, 2003.

\end{thebibliography}

\end{document}